\documentclass[prl,twocolumn,preprintnumbers,superscriptaddress,showpacs]{revtex4-1}

\usepackage{graphicx}
\usepackage{bm}
\usepackage{amsmath}
\usepackage{amssymb}
\usepackage{slashed}
\usepackage[colorlinks,pdfstartview=FitH]{hyperref}
\hypersetup{linkcolor=blue,citecolor=blue,filecolor=black,urlcolor=blue}

\begin{document}
\title{Macroscopic transports in a rotational system with an electromagnetic field}
\author{Gaoqing Cao}
\affiliation{School of Physics and Astronomy, Sun Yat-sen University, Zhuhai 519088, China}
\date{\today}

\begin{abstract}
In this work, we explore macroscopic transport phenomena associated with a rotational system in the presence of an external orthogonal electromagnetic field. Simply based on the lowest Landau level approximation, we derive nontrivial expressions for chiral density and various currents consistently by adopting small angular velocity expansion or Kubo formula. While the generation of anomalous electric current is due to the pseudo gauge field effect of the spin-rotation coupling, the chiral density and current can be simply explained with the help of Lorentz boosts. Finally, Lorentz covariant forms can be obtained by unifying our results and the magnetovorticity effect.
\end{abstract}

\maketitle
\section{Introduction}
In 1969, Adler~\cite{Adler:1969gk}, Bell and Jackiw~\cite{Bell:1969ts} discovered that in $3+1$ dimensional quantum electrodynamics (QED), although the axial symmetry at tree level holds, the one-loop triangle diagram gives rise to a non-conserved axial current $J_5^\mu$,
\begin{eqnarray}
\label{QEDanom}
\partial_\mu J^\mu_5=-{e^2\over16\pi^2}\epsilon^{\alpha\beta\mu\nu}F_{\alpha\beta}F_{\mu\nu},
\end{eqnarray}
where $F_{\mu\nu}=\partial_\mu A_\nu-\partial_\nu A_\mu$ is the Maxwell tensor of electromagnetic (EM) field.
This result contradicts the naive expectation from Noether's theorem and thus was referred to as a ``chiral anomaly" or ``triangle anomaly". The extension of the above chiral anomaly to quantum chromodynamics (QCD) is straightforward which reads
\begin{eqnarray}
\label{QCDanom}
\partial_\mu J^\mu_5=-{g^2N_{\rm f}\over32\pi^2}\epsilon^{\alpha\beta\mu\nu}F_{\alpha\beta}^cF_{\mu\nu}^c,
\end{eqnarray}
where $F_{\mu\nu}^c$ is the non-abelian field tensor and $N_{\rm f}$ is the number of quark flavors.

Chiral anomalies play prominent roles in modern physics and underlie a number of important phenomena like the large decay rate of $\pi^0\rightarrow\gamma\gamma$~\cite{Bell:1969ts}, the large mass splitting between $\eta$ and $\eta'$ mesons~\cite{Witten:1979vv}, and baryogenesis in the early Universe~\cite{Farrar:1993sp}. The interplay of the QED and QCD contributions to chiral anomaly provides a feasible means to detect the topological fluctuations of QCD in heavy-ion collision experiments through the so-called chiral magnetic effect (CME)~\cite{Kharzeev:2007jp,Fukushima:2008xe}. Owing to Eq.~(\ref{QCDanom}), finite chiral imbalance is generated in QCD system and can be characterized by chiral chemical potential  $\mu_5$; then a magnetic field ${\bf B}$ induces a macroscopic electric current $\bf J$ along its direction according to Eq.~(\ref{QEDanom}). Similar to the ordinary conductors with finite electric field, the CME of quarks can be presented as
\begin{eqnarray}
\label{cmecurrent}
{\bf J}=\sigma_5{\bf B},\ \sigma_5=\frac{N_c}{2\pi^2}\sum_{\rm f}q_{\rm f}^2\mu_5
\end{eqnarray}
with the summation over all relevant quark flavors. It is fantastic that the CME current is non-dissipative due to the time-reversal-even nature of the CME conductivity $\sigma_5$  and was justified according to the absence of drag force acting on an impurity put into the flow~\cite{Rajagopal:2015roa,Stephanov:2015roa,Sadofyev:2015tmb}. Recently, the CME has been experimentally realized in Weyl semimetal ZrTe5  in condensed matter physics~\cite{Li:2014bha} and the isobar collision experiments are being carried out at the Relativistic Heavy Ion Collider to detect the CME~\cite{Shi:2019wzi,Kharzeev:2020jxw,STAR:2020crk}.In the past two decades, other macroscopic transports relevant to chiral anomaly are also intensively studied from high-energy nuclear physics to condensed matter physics, for example, the chiral vortical effect~\cite{Erdmenger:2008rm,Son:2009tf,Banerjee:2008th}, the chiral separation effect~\cite{Son:2004tq,Metlitski:2005pr}, the chiral electric separation effect~\cite{Huang:2013iia}, anomalous magnetovorticity~\cite{Hattori:2016njk}, and so on; see reviews Refs.~\cite{Liao:2014ava,Kharzeev:2015kna,Huang:2015oca} for more details.

\begin{figure}[!htb]
	\begin{center}
		\includegraphics[width=6cm]{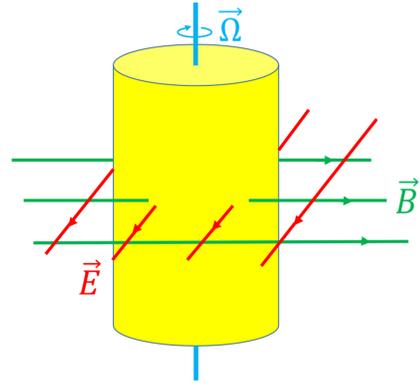}
		\caption{A rotating cylindrical system in the presence of an external electromagnetic field, where the electric field ${\bf E}$, magnetic field ${\bf B}$ and angular velocity $\mathbf{\Omega}$ span a Cartesian coordinate system.}\label{EBO}
	\end{center}
\end{figure}

In this work, we propose another special circumstance where new macroscopic transport phenomena would emerge, see the experimental setup for massless fermions in Fig.\ref{EBO}. 
With the discoveries of strong electromagnetic (EM) field~\cite{Voronyuk:2011jd,Bzdak:2011yy,Deng:2012pc} and longitudinal local vorticity~\cite{Becattini:2017gcx,Becattini:2019ntv,Xia:2019fjf,Xia:2018tes,Niida:2018hfw,Adam:2019srw} in heavy ion collisions, such setup can be relevant to relativistic heavy ion experiments in high energy colliders. As the electric field is usually screened and the rotation is limited by the system size due to causality, we set $|e{\bf B}|\gg |e{\bf E}|, \boldsymbol{\Omega}^2$ here. Then, under the lowest Landau level (LLL) approximation, we apply small angular velocity $\Omega$ expansion or Kubo formula, and obtain consistent results for the macroscopic density and currents:
\begin{eqnarray}\label{CEMVE}
n_5={|q|({\bf E}\times\boldsymbol{\Omega})\cdot{\bf B}\over4\pi^2 I_1},\;
{\bf J}_5=-{|q|{\bf E}^2\over4\pi^2I_1}\boldsymbol{\Omega},\;
{\bf J}={q{\bf E}\times\boldsymbol{\Omega}\over4\pi^2}
\end{eqnarray}
with the first Lorentz invariant for EM field $I_1=\left({\bf B}^2-{\bf E}^2\right)^{1/2}$. 

\begin{figure}[!htb]
	\begin{center}
		\includegraphics[width=6cm]{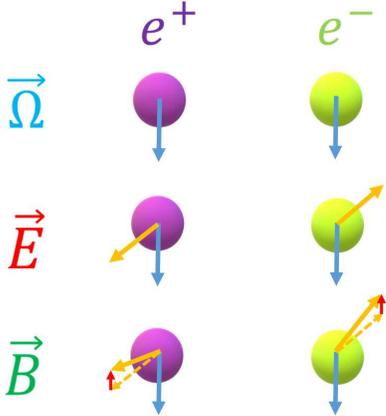}
		\caption{An illumination of chirality generation: the blue and yellow arrows denote the preferred spin and momentum directions, respectively; and the red arrows stand for the momentum shifts induced by the magnetic field.}\label{chirality}
	\end{center}
\end{figure}

According to these equations, we expect to detect vector current ${\bf J}$ along ${\bf B}$ and axial current ${\bf J}_5$ along $\mathbf{\Omega}$. Though the generation of ${\bf J}$ can be easily understood by following CME as chiral imbalance $n_5$ is simultaneously generated in our setup, it doesn't depend on the presence of ${\bf B}$ at all. Note that ${\bf J}$ is not induced by the classical Hall-like effect since the contribution is soly from the spin-rotation coupling. Furthermore, the chiral imbalance generation mechanism are illuminated in Fig.\ref{chirality} and can be simply explained in this way: the Poynting vector ${\bf S}\equiv{\bf E\times B}$ pushes both particles and antiparticles parallel to itself; then the charge-blind spin polarization along $\mathbf{\Omega}$, antiparallel to ${\bf S}$ here, favors the helicity $-1$ over $+1$ for both particles and antiparticles. 
For later convenience, these macroscopic transport phenomena will be denoted as "chiral electromagnetic vortical effect" (CEMVE) all together with the generation of ${\bf J}$ especially named as "chiral electric vortical effect" (CEVE). To the best of our knowledge, the CEMVE phenomena are completely novel, thus the CEVE constitutes a new member of the chiral anomaly transport family. 

\section{Small rotation assumption}
Under the circumstance of a global rotation and a constant EM field as illuminated in Fig.\ref{EBO},  the Lagrangian of a free fermion system is given by
\begin{eqnarray}\label{Lcv}
{\cal L}=\bar\psi\left[i\gamma^\mu(D_\mu+\Gamma_\mu)-m\right]\psi.
\end{eqnarray}
Here, $\psi$ is chosen to be a one-flavor fermion field for simplicity and the fermion mass $m$ will be set to zero for the study of chiral anomalies. For the covariant derivative $D_\mu=\partial_\mu+iqA_\mu$, we choose the vector potential $A_\mu=(iEx_1,0,-Bx_1,0)$ in Euclidean space without lose of generality. Then, the orthogonal rotation effect can be introduced by altering the flat-space metric $\eta_{\mu\nu}={\rm diag}(1,-1,-1,-1)$ to the curved-space one~\cite{Parker2009}:
\begin{eqnarray}\label{gcv}
g_{\mu\nu}=\left(\begin{array}{cccc}
1-(x_1^2+x_3^2)\Omega^2&-x_3\Omega&0&x_1\Omega\\
-x_3\Omega&-1&0&0\\
0&0&-1&0\\
x_1\Omega&0&0&-1
\end{array}\right).
\end{eqnarray}
And the spin-rotation coupling is presented by the affine connection $\Gamma_\mu$ which is defined in terms of the spin connections $\omega_{\mu ij}$ or vierbeins $e_{i}^\mu$ as~\cite{Parker2009}
\begin{eqnarray}\label{gammacv}
\Gamma_\mu=-{i\over4}\omega_{\mu ij}\sigma^{ij},~\omega_{\mu ij}=g_{\alpha\beta}e^\alpha_{\rm i}\left(\partial_\mu e^\beta_{\rm j}+\Gamma^\beta_{\mu\nu}e^\nu_{\rm j}\right)
\end{eqnarray}
with the Greek and Latin indices for coordinate and tangent spaces, respectively. Note that the Christoffel symbol $\Gamma^\beta_{\mu\nu}$ is related to the derivatives of $g_{\mu\nu}$ with respect to the coordinates. Substituting Eqs.(\ref{gcv}) and (\ref{gammacv}) into Eq.(\ref{Lcv}), we eventually get a flat-space Lagrangian:
\begin{eqnarray}\label{Lagarangion}
\!\!\!\!\!{\cal L}=\bar\psi\left(i\slashed{{D}}-\hat{\Omega}_2\right)\psi,\;\hat{\Omega}_2\!\equiv\! i\gamma^0\Omega \left(\!x_3{D}_1\!-\!x_1{D}_3\!+\!{i\over2}\sigma^{31}\!\right)
\end{eqnarray}
with the spin operator $\sigma^{31}\equiv{i\over2}\left[\gamma^3,\gamma^1\right]$.

For the setup of ${\bf E}, {\bf B}$ and $\mathbf{\Omega}$ in Fig.\ref{EBO}, it is hard to solve the explicit form of fermion propagator, a basic computing element in quantum field theory. However, the quantum EM effect has already been well studied by J.S. Schwinger in 1951 and the fermion propagator ${G}(E,B)$ was generally given in the form of proper time integral~\cite{Schwinger:1951nm}. To facilitate our discussions, we can make use of that powerful result by assuming the rotation is relatively weak, that is, $\Omega\ll I_1^{1/2}$. Then, by expanding around small ${\Omega}$ to the first order, the full fermion propagator in Eq.(\ref{Lagarangion}) becomes approximately
\begin{eqnarray}
{G}(E,B,\Omega)\!\equiv\!{1\over i\slashed{{D}}-\hat{\Omega}_2}\!\approx\!{G}(E,B)\left[1+\hat{\Omega}_2{G}(E,B)\right].
\end{eqnarray}
 Only the second term is responsible for the emergence of macroscopic density and currents in the system.
 
\subsection{Lowest Landau level}
In order to avoid renormalization ambiguity, we just focus on the lowest Landau level which was found to be the main origin of chiral anomaly~\cite{Kharzeev:2007jp,Hattori:2016njk}. For pure magnetic field, we have $B=I_1$ and the LLL fermion propagator can be given in Euclidean space as~\cite{Miransky:2015ava}
\begin{eqnarray}\label{SM}
G_{\rm LLL}^{\rm M}(x,x')&=&e^{i\Phi^{\rm M}}{|qI_1|\over2\pi}e^{-{|qI_1|\over4}\left(\delta x_1^2+\delta x_2^2\right)}G_{1+1}(\delta x_3,\delta x_4)\nonumber\\
&&{1+i\,{\cal S}_{\rm q}\gamma^1\gamma^2\over2},\;\delta x=x-x',
\end{eqnarray}
where ${\cal S}_{\rm q}$ is sign of $q$, the Schwinger phase is defined as $$\Phi\equiv q\int_{x'}^xdy^\mu\left[A_\mu(y)+{1\over2}F_{\mu\nu}(y-x')^\nu\right]$$ and $G_{1+1}(\delta x_3,\delta x_4)$ is the fermion propagator in $1+1$ dimensions. Then, for the general case with ${\bf B}\bot{\bf E}$, the fermion propagator can be obtained from Eq.\eqref{SM} by taking a Lorentz boost along ${\bf S}$, that is,
\begin{eqnarray}
\Lambda_L=\left(\begin{array}{cccc}
\beta&0&i\epsilon&0\\0&1&0&0\\-i\epsilon&0&\beta&0\\0&0&0&1
\end{array}\right)
\end{eqnarray}
with $\beta={B/I_1}$ and $\epsilon={E/I_1}$.
So we have $G_{\rm LLL}^{\rm EM}(x,x')=e^{i\Phi^{\rm EM}}\tilde{G}_{\rm LLL}^{\rm EM}(\delta x)$ where the translation invariant part is given by
\begin{eqnarray}\label{SEM}
\!\!\!\!\tilde{G}_{\rm LLL}^{\rm EM}(x)\!\equiv\!{|qI_1|\over2\pi}e^{-{|qI_1|\over4}\left(\!x_1^2\!+\tilde{x}_2^2\!\right)}\tilde{G}_{1+1}(x_3,\tilde{x}_4){1\!+\!i\,{\cal S}_{\rm q}\gamma^1\tilde{\gamma}^2\over2}
\end{eqnarray}
with the Lorentz boosted coordinates and gamma matrices: 
\begin{eqnarray}
&&\tilde{x}_2=\beta x_2-i\epsilon x_4,\ \tilde{x}_4=\beta x_4+i\epsilon x_2,\nonumber\\
&& \tilde{\gamma}^2=\beta\gamma^2-i\epsilon\gamma^4,\ \tilde{\gamma}^4=\beta\gamma^4+i\epsilon\gamma^2.
\end{eqnarray}
For future convenience, we take Fourier transformation of $\tilde{G}_{\rm LLL}^{\rm EM}(x)$ and get the effective propagator in energy-momentum space as
\begin{eqnarray}
\tilde{G}_{\rm LLL}^{\rm EM}(p)
&=&{-i\,e^{-{(p_1^2+\tilde{p}_2^2)/|qI_1|}}\over \gamma^3p_3+\tilde{\gamma}^4\tilde{p}_4}\left({1+i\,{\cal S}_{\rm q}\gamma^1\tilde{\gamma}^2}\right).
\end{eqnarray}

Eventually, the expectation value of chiral density can be evaluated as:
\begin{eqnarray}
n_5&\equiv&{1\over V_{3+1}}\int{d^4x}\langle\bar\psi(x)\gamma^0\gamma^5\psi(x)\rangle={-1\over V_{3+1}}{\rm Tr}\,\gamma^0\gamma^5{S}(x,x')\nonumber\\
&=&-{\Omega}\int{d^4x}\;{\rm tr}\,\gamma^0\gamma^5\tilde{G}_{\rm LLL}^{\rm EM}(-x)\gamma^4\left[x_3\left(\partial_1-{iI_1}{\tilde{x}_2\over2}\right)\right.\nonumber\\
&&\ \ \ \ \ \ \ \  \ \ \ \ \ \ \ \ \ \left.-x_1\partial_3+{1\over2}\gamma^1\gamma^3\right]\tilde{G}_{\rm LLL}^{\rm EM}(x).
\end{eqnarray}
After checking the odd-evenness of the integrand with respect to the coordinates, we find that only the spin-rotation coupling contributes thus
\begin{eqnarray}
n_5&=&\!\!-{\Omega\over2}\int{d^4x}\;{\rm tr}\,\left[\gamma^0\gamma^5\tilde{G}_{\rm LLL}^{\rm EM}(-x)\gamma^4\gamma^1\gamma^3\tilde{G}_{\rm LLL}^{\rm EM}(x)\right]\nonumber\\
&=&\!\!-i{\Omega}{|qI_1|\over16\pi}\int{dp_3dp_4\over{(2\pi)^2}}\;{\rm tr}\,\left[\left(\beta\tilde{\gamma}^4\!-\!i\epsilon\tilde{\gamma}^2\right)\gamma^5{1\!+\!i\,{\cal S}_{\rm q}\gamma^1\tilde{\gamma}^2\over \gamma^3p_3\!+\!\tilde{\gamma}^4p_4}\right.\nonumber\\
&&\ \ \ \ \ \ \ \ \ \ \ \ \ \ \left.\left(\beta\tilde{\gamma}^4\!-\!i\epsilon\tilde{\gamma}^2\right)\gamma^1\gamma^3{1\!+\!i\,{\cal S}_{\rm q}\gamma^1\tilde{\gamma}^2\over \gamma^3p_3\!+\!\tilde{\gamma}^4p_4}\right]\nonumber\\
&=&\!\!-{\Omega}{|q|EB\over4\pi I_1}\!\lim_{q_0\rightarrow0}\!\lim_{q_3\rightarrow0}\!i\langle J^3(q)J^3(-q)\rangle_{\rm 2D}\!=\!{\Omega}{|q|EB\over4\pi^2 I_1}.\label{n51}
\end{eqnarray}
Here, we have made use of a new gamma matrix representation: $\gamma^1, \tilde{\gamma}^2, \gamma^3$ and $\tilde{\gamma}^4$, and the dimensional regularization together with gauge invariant condition or equivalently bosonization rule~\cite{Smilga:1992hx,Fukushima:2011jc,Fukushima:2015wck} is adopted to arrive at the final result in the last step. Together with that, a chiral current along $\mathbf{\Omega}$ is also generated, that is,
\begin{eqnarray}
J_{5}^{2}&=&-{\Omega\over2}\int{d^4x}\;{\rm tr}\,\left[\gamma^2\gamma^5\tilde{G}_{\rm LLL}^{\rm EM}(-x)\gamma^4\gamma^1\gamma^3\tilde{G}_{\rm LLL}^{\rm EM}(x)\right]\nonumber\\
&=&-{\Omega}{|q|E\epsilon\over4\pi^2}=-{\Omega}{|q|E^2\over4\pi^2 I_1}.
\end{eqnarray}
Actually, the internal steps can be simply understood in this way: In the pure magnetic field frame, the spin-rotation coupling gives rise to an effective chiral chemical potential term ${\epsilon\,\Omega\over2}\tilde{\gamma}^0\gamma^5$ and the corresponding chiral density is $\tilde{n}_5={|q|E\Omega/(4\pi^2)}$. Then, we get $n_5=\beta\tilde{n}_5$ and $J_{5}^{2}=-\epsilon\tilde{n}_5$ by boosting back to the laboratory frame with ${\bf B}\bot{\bf E}$. Here, since the generations of chiral density and current only involve Lorentz boosts but no triangle diagrams, they are not effects due to chiral anomaly at all.

Similarly, the vector current along ${\bf B}$ can be evaluated as:
\begin{eqnarray}
J^3&=&-{\Omega\over2}\int{d^4x}\;{\rm tr}\,\left[\gamma^3\tilde{G}_{\rm LLL}^{\rm EM}(-x)\gamma^4\gamma^1\gamma^3\tilde{G}_{\rm LLL}^{\rm EM}(x)\right]\nonumber\\
&=&{\Omega}{qE\over4\pi}\!\lim_{q_0\rightarrow0}\!\lim_{q_3\rightarrow0}(-i\langle J^3(q)J^3(-q)\rangle_{2D}){=}{\Omega}{qE\over4\pi^2}.\label{J31}
\end{eqnarray}
Note that while the axial current is blind to the sign of the electric charge, the vector current is not. In the following, we demonstrate that such effect is from chiral anomaly. For that purpose, after suppressing the orbital-rotation coupling as before, we can rewrite the Lagrangian Eq.\eqref{Lagarangion} as a sum of right- and left-handed parts by adopting Weyl representation, that is,
\begin{eqnarray}\label{L_H}
{\cal L}=\sum_{\rm H=R,L}\bar\psi_H\left(i\sigma^{\mu}_{\rm H} D_{\mu H}+{\cal S}_{\rm H}\sigma^{2}_{\rm H}{\Omega\over2}\right)\psi_{\rm H}.
\end{eqnarray}
Here, we've defined $\sigma^{\mu}_{\rm H}=(-i\,I_{2\times2},{\cal S}_{\rm H}\boldsymbol{\sigma})$ with $\boldsymbol{\sigma}$ Pauli matrices and ${\cal S}_{\rm H}=\pm$ for right- and left-handed fermions, respectively. From the new expression, we can immediately recognize the correspondence: $q{\bf B}\leftrightarrow -{{\cal S}_{\rm H}\over2}\nabla\times\mathbf{\Omega}$, that is, the spin-rotation coupling plays a role of pseudo gauge field~\cite{Guinea2010,Liu2013}. Actually, the chiral anomaly in QED can be alternatively presented as
\begin{eqnarray}
\partial_\mu J^\mu_{\rm H}={\cal S}_{\rm H}{q^2\over4\pi^2}\bf{E\cdot B},
\end{eqnarray}
then the correspondence implies
\begin{eqnarray}
\partial_\mu J^\mu_{\rm H}=-{q\over8\pi^2}{\bf E\cdot (\nabla\times\mathbf{\Omega})}={q\over8\pi^2}\nabla\cdot(\bf{E}\times\mathbf{\Omega})
\end{eqnarray}
in a rotational system with pure electric field. Thus, ${\bf J}=\sum_{\rm H=R,L}{\bf J}_{\rm H}$ exactly gives the vector current shown in Eq.(\ref{CEMVE}), which actually does not request for the presence of magnetic field at all.

\section{Kubo formula}
As we know, Kubo formula is very useful to derive the transport coefficients for the densities and currents in quantum field theory (QFT)~\cite{Hattori:2016njk,Amado:2011zx}. For the chosen form of gauge field: $A_\mu=(a_{\rm 0}(x_1),0,a_{\rm 2}(x_1),0)$, we set the metric as $g_{\mu\nu}=\eta_{\mu\nu}+h_{\mu\nu}$ with the strains $h_{\rm 02},h_{\rm 03},h_{\rm 12}$ and $h_{\rm 13}$ nonzero and $x_1$ dependent. Then up to the first order of spatial derivatives, sources and velocity, the constitutive relations are given by~\cite{Amado:2011zx}
\begin{eqnarray}
T^{\rm 0m}&=&(\epsilon+P)\upsilon^{\rm m}+P\,h_{\rm 0m},\nonumber\\
J^{\rm m}&=&n\,\upsilon^{\rm m}\!+\!\epsilon^{\rm mn}\xi_{V}(\partial_{\rm 1}h_{\rm 0n}\!+\!\partial_{\rm 1}\upsilon^{\rm n})\!+\!\xi_B\epsilon^{\rm mn}\partial_{\rm 1}a_{\rm n}
\end{eqnarray}
in the static limit, where $\epsilon^{\rm mn}$ is the Levi-Civita symbol with the indices $m,n=2,3$. By following the discussions in Ref.~\cite{Amado:2011zx}, we can easily check that Kubo formula also applies here.

Take $n_5$ and ${\bf J}$ for example. Recalling their properties under charge conjugate, parity and time reversal (CPT) transformations and Lorentz boost, they should take the following forms:
\begin{eqnarray}
n_5&=&\lambda_5\epsilon^{\rm ijk}\hat{E}_{\rm i}\hat{B}_{\rm j}\Omega_{\rm k}={\lambda_5\over2}\epsilon^{\rm ijk}\epsilon^{\rm klm}\hat{E}_{\rm i}\hat{B}_{\rm j}(iq_{\rm l})g_{\rm 0m},\label{KFn5}\\
J^i&=&\lambda_{\rm i}\epsilon^{\rm ijk}\hat{E}_{\rm j}\Omega_{\rm k}={\lambda_{\rm i}\over2}\epsilon^{\rm ijk}\epsilon^{\rm klm}\hat{E}_{\rm j}(iq_{\rm l})g_{\rm 0m}.\label{KFJi}
\end{eqnarray}
Here, $\hat{E}$ and $\hat{B}$ are unit electric and magnetic vectors, and the angular velocity is now given by ${\boldsymbol \Omega}\equiv{1\over2}\nabla\times{\boldsymbol \upsilon}$ with the velocity ${\upsilon_{\rm m}}$ corresponding to the gravitomagnetic potential $g_{\rm 0m}$~\cite{Amado:2011zx}. Then, we set ${\boldsymbol \upsilon}=\upsilon_3(x_1)\hat{e}_3$ for simplicity and find with the help of the property $\partial_{g_{03}}=T^{03}$:
\begin{eqnarray}
\!\!\!\!n_5&=&{\lambda_5\over2}(iq_{1})g_{03},\,\ \ \lambda_5=\lim_{{q_1}\rightarrow0}{2\over iq_{1}}
\lim_{q_{\mu\neq1}\rightarrow0} G^{5,03}_{\rm R}(q);\label{L5}\\
\!\!\!\!J^3&=&-{\lambda_3\over2}(iq_{1})g_{03},\,\lambda_3=\lim_{{q_1}\rightarrow0}{2i\over q_{1}}
\lim_{q_{\mu\neq1}\rightarrow0} G^{3,03}_{\rm R}(q);\label{L3}
\end{eqnarray}
where the retarded Green's functions are Fourier correspondences to the following ones in coordinate space:
\begin{eqnarray}
G^{5,03}_{\rm R}(x-x')&\equiv&\langle n_5(x)T^{03}(x')\rangle\theta(t-t'),\\
G^{3,03}_{\rm R}(x-x')&\equiv&\langle J^3(x)T^{03}(x')\rangle\theta(t-t').
\end{eqnarray}

In QFT, the energy-momentum tensor is given by
\begin{eqnarray}
T^{03}(x)={i\over2}\bar{\psi}(x)(\gamma^0D^3+\gamma^3D^0)\psi(x),
\end{eqnarray}
hence $G^{5,03}_{\rm R}(q)$ can be evaluated in the limits $q_{\mu\neq1}\rightarrow0$ and $q_1\sim 0$ as
\begin{eqnarray}
G^{5,03}_{\rm R}\!\!&=&\! iq_1{{\cal S}_{\rm q}\epsilon \over4}\!\int\!\!{d^4p\over{(2\pi)^4}}{\rm Tr}\Big[\gamma^0\gamma^5\tilde{G}_{\rm LLL}^{\rm EM}(p\!+\!q)\gamma^3\tilde{G}_{\rm LLL}^{\rm EM}(p)\Big]\nonumber\\
\!\!&=&\!iq_1\!{|q|EB \over8\pi I_1}\!\!\lim_{q_0\rightarrow0}\!\lim_{q_3\rightarrow0}\! i\langle J^3(q)J^3(-q)\rangle_{2D}\!=\!\!-iq_1\!{|q|EB \over8\pi^2 I_1}.\nonumber\\
\end{eqnarray}
Note that such $q_1$ linear feature originates from operating $D^0$, involved in $T^{03}(x)$, to the Schwinger phase~\cite{Hattori:2016njk}, and all the other contributions automatically vanish in the limits considered. In a similar way, the other retarded Green's function can be straightforwardly calculated as
\begin{eqnarray}
\!\!\!\!\!\!G^{3,03}_{\rm R}\!\!&=&\! iq_1{{\cal S}_{\rm q}\epsilon \over4}\!\int\!\!{d^4p\over{(2\pi)^4}}{\rm Tr}\Big[\gamma^3\tilde{G}_{\rm LLL}^{\rm EM}(p\!+\!q)\gamma^3\tilde{G}_{\rm LLL}^{\rm EM}(p)\Big]\nonumber\\
\!\!\!\!\!\!\!\!&=&\!iq_1{qE\over8\pi}\!\!\lim_{q_0\rightarrow0}\!\lim_{q_3\rightarrow0}\! i\langle J^3(q)J^3(-q)\rangle_{2D}\!=\!-iq_1{qE \over8\pi^2}.\label{G303}
\end{eqnarray}

Eventually, Eqs.\eqref{L5} and \eqref{L3} give
\begin{eqnarray}
\lambda_5=-{|q|EB \over4\pi^2 I_1},\ \ \ \ \lambda_3={qE \over4\pi^2},
\end{eqnarray}
which are consistent the results derived in the previous section. Substituting the coefficients into Eqs.\eqref{KFn5} and \eqref{KFJi}, $n_5$ and ${\bf J}$ can be represented in the forms of vector algorithm, see Eq.(\ref{CEMVE}).

\section{Summary and discussions}
In this letter, we found novel macroscopic transport phenomena under the circumstance of orthogonal rotation and EM field. By adopting fermion propagator in the lowest Landau level approximation, we managed to derive consistent results for the density and current generations Eq.(\ref{CEMVE}) with small rotation assumption or Kubo formula. If we focus on the spin-rotation coupling, the results actually persist in full Landau level evaluations~\cite{supp}. Recalling the magnetovorticity effect for the case $\boldsymbol{\Omega}\parallel{\bf B}$~\cite{Hattori:2016njk}, the four-(axial)vector currents $J^\mu$ and $J^\mu_5$ can be generally put in Lorentz covariant forms as
\begin{eqnarray}
J^\mu={q\over4\pi^2}\tilde{F}^{\mu\nu}\Omega_\nu,\ \
J^\mu_5={|q|\over4\pi^2I_1}\tilde{F}^{\mu\nu}\tilde{F}_{\nu\rho}\Omega^\rho
\end{eqnarray}
with the dual strength tensor $\tilde{F}^{\mu\nu}=\epsilon^{\mu\nu\rho\sigma}F_{\rho\sigma}/2$. Here, we've compensated a temporal component for the angular velocity: $\Omega^0=\gamma\,\boldsymbol{\Omega}\cdot {\bf V}$ with $\gamma$ the Lorentz factor and ${\bf V}$ the velocity of the observer's frame.

The divergence of $J_5^\mu$ at $I_1=0$ seems weird. Actually, the point is that we didn't take into account the boundary effect~\cite{Ebihara:2016fwa,Liu:2017spl,Cao:2020pmm}, which is necessary for a rotational system to satisfy causality. A finite boundary will induce a mass gap $\delta m$ to the fermions in the bulk~\cite{Ebihara:2016fwa,Liu:2017spl,Cao:2020pmm}; then take chiral density for example, the final expression alters to
\begin{eqnarray}
n_5(z)=-\Omega {|q|EB\over4\pi^2I_1}\left[1+2x(\psi_0(z)-\ln z)\right]
\end{eqnarray}
with the auxiliary variable $z=\delta m^2/(2|qI_1|)$ and $\psi_0(z)$ the digamma function. In two opposite limits of $z$, we find
\begin{eqnarray}
\lim_{z\rightarrow0}n_5(z)=\Omega {|q|EB\over4\pi^2I_1},\ 
\lim_{z\rightarrow\infty}n_5(z)=\Omega {q^2EB\over12\pi^2\delta m^2}.
\end{eqnarray}
Thus, the divergence is safely avoided but the chiral current $J_5^\mu$ indeed can be greatly enhanced by reducing $I_1$.

For QCD, Eq.(\ref{CEMVE}) should be modified by taking into account the relevant flavor and color degrees of freedom. With the CME experiments going on in RHIC, the CEMVE can also play a role in heavy ion collisions, where the temperature effect might be important to the non-anomalous axial current $J_5^\mu$. In the future, the effects of temperature, orbital-rotation coupling and electric field dominance will be explored in more detail by adopting the full fermion propagator in orthogonal EM field~\cite{Schwinger:1951nm}. 

\emph{Acknowledgments}---GC thanks Xingyu Guo's helpful comment when visiting Tsinghua University and appreciates  Xu-guang Huang, Hao-Lei Chen, and Kazuya Mameda for their checks and useful discussions. G.C. is supported by the National Natural Science Foundation of China with Grant No. 11805290.

\end{document}


\title{Supplemental material: the derivations with full Landau levels}

\author{Gaoqing Cao}
\affiliation{School of Physics and Astronomy, Sun Yat-sen University, Zhuhai 519088, China}
\date{\today}


\maketitle
Here, we focus on the spin-rotation coupling term ${\Omega\over2}\gamma^4\gamma^1\gamma^3$ and check Eq.(4) in the main text with full Landau levels evaluations. The fermion propagator with full effect of external electromagnetic (EM) field had been given in a very general and compact form in 1951 by Schwinger~\cite{Schwinger:1951nm}. In the case of constant orthogonal EM field: $\bf{B\bot E}$, the full propagator can be formally presented as ${\cal S}(x,x')\equiv e^{i\Phi^{\rm M}} \hat{\cal S}({\delta x})$ with $\Phi^{\rm M}$ and $\delta x$ defined in Eq.(10) of the main text. To our main concern, the Schwinger phase $e^{i\Phi^{\rm M}}$ cancels out as in Eq.(16) of the main text, so we can transform the effective propagator $\hat{\cal S}({\delta x})$ to energy-momentum space for later convenience and get
\begin{eqnarray}
\hat{\cal S}({p})
&=&i\int_0^\infty {ds}~
\exp\left[-is\left(m^2+{p}_3^2+(\beta{p}_4+i\epsilon{p}_2)^2\right)-i\varphi(s)\left({p}_1^2+(\beta{p}_2-i\epsilon{p}_4)^2\right)\right]~\Big[-\gamma^3{p}_3+m-\gamma^4\Big({p}_4-iqE\varphi(s){p}_1\Big)
\nonumber\\
&&
-\gamma^1\Big({p}_1+iqE\varphi(s){p}_4-qB\varphi(s){p}_2\Big)
-\gamma^2\Big({p}_2+qB\varphi(s){p}_1\Big)\Big]\Big[1+i{qE}\varphi(s)\gamma^4\gamma^1+{qB}\varphi(s)\gamma^1\gamma^2\Big],
\end{eqnarray}
where the auxiliary function is $\varphi(s)=\tan\left(qI_1s\right)/\left(qI_1\right)$. 

Take the chiral density $n_5$ for example, we can evaluate it by following Eq.(16) of the main text as
\begin{eqnarray}
n_5&=&-{\Omega\over2}{\rm Tr}\;\gamma^0\gamma^5\hat{\cal S}({p})\gamma^4\gamma^1\gamma^3\hat{\cal S}({p})\nonumber\\
&=&{4\Omega\,qE\,qB\over (qI_1)^2}\int {\rm d}\,s\;{\rm d}\,t~\int{{\rm d}^4P\over(2\pi)^4}\exp\left[-i(s+t)\left(m^2+{P}_3^2+P_4^2\right)-i[\varphi(s)+\varphi(t)]\left({P}_1^2+P_2^2\right)\right]\times\nonumber\\
&&\left[(m^2-P_3^2)\tan(qI_1s)\tan(qI_1t)-P_4^2+(1+\tan^2(qI_1s))(1+\tan^2(qI_1t))P_2^2\right],
\end{eqnarray}
where we have taken variable transformations: $$\beta{p}_4+i\epsilon{p}_2\rightarrow P_4,\ \beta{p}_2-i\epsilon{p}_4\rightarrow P_2,\ p_1\rightarrow P_1,\ p_3\rightarrow P_3$$ and gotten rid of vanishing terms with linear $P$. Then, by taking proper-time transformations $s\rightarrow-i\,s$ and $t\rightarrow-i\,t$ and completing the energy-momentum integrals, we obtain 
\begin{eqnarray}
n_5
&=&{\Omega\,qE\,qB\over 4\pi^2qI_1}\int {\rm d}\,s\;{\rm d}\,t{e^{-(s+t)m^2}\over s+t}{m^2\tanh(qI_1s)\tanh(qI_1t)+{1\over 2(s+t)}(1-\tanh(qI_1s)\tanh(qI_1t))-{(1-\tanh^2(qI_1s))(1-\tanh^2(qI_1t))\over2(\tanh(qI_1s)+\tanh(qI_1t))/(qI_1)}\over\tanh(qI_1s)+\tanh(qI_1t)}.
\end{eqnarray}
Finally, we take variable transformations $s\rightarrow s{1+u\over 2}$ and $t\rightarrow s{1-u\over 2}$, carry out the integral over $u\in(-1,1)$, and find
\begin{eqnarray}
n_5
&=&{\Omega\,qE\,qB\over 8\pi^2qI_1}\int_0^\infty {\rm d}\,s~e^{-s\,m^2}\left[m^2\left({1\over\tanh(qI_1s)}-{1\over qI_1s}\right)+{1\over qI_1s^2}-{qI_1\over \sinh^2(qI_1s)}\right]\nonumber\\
&=&{\Omega\,qE\,qB\over 4\pi^2qI_1}\int_0^\infty {\rm d}\,s~e^{-s\,m^2}m^2\left({1\over\tanh(qI_1s)}-{1\over qI_1s}\right)\stackrel{m\rightarrow0}{=}{\Omega\,|q|E\,B\over 4\pi^2I_1},
\end{eqnarray}
which is again the same as that in Eq.(4) of the main text.

Similarly, by omitting the internal steps up to the energy-momentum integrals, the axial current can be evaluated as:
\begin{eqnarray}
J_5^2&=&-{\Omega\over2}{\rm Tr}\;\gamma^2\gamma^5\hat{\cal S}({p})\gamma^4\gamma^1\gamma^3\hat{\cal S}({p})\nonumber\\
&=&{qI_1\Omega\over8\pi^2}\int {\rm d}\,s\;{\rm d}\,t{e^{-(s+t)m^2}\over s+t}{m^2\left[1-{B^2+E^2\over I_1^2}\tanh(qI_1s)\tanh(qI_1t)\right]-{B^2\over I_1^2}{1\over s+t}(1-\tanh(qI_1s)\tanh(qI_1t))+{E^2\over I_1^2}{(1-\tanh^2(qI_1s))(1-\tanh^2(qI_1t))\over(\tanh(qI_1s)+\tanh(qI_1t))/(qI_1)}\over\tanh(qI_1s)+\tanh(qI_1t)}\nonumber\\
&=&{qI_1\Omega\over16\pi^2}\int {\rm d}\,s~e^{-s\,m^2}\left\{m^2\left[\left({1\over\tanh(qI_1s)}+{1\over qI_1s}\right)-{B^2+E^2\over I_1^2}\left({1\over\tanh(qI_1s)}-{1\over qI_1s}\right)\right]-{B^2\over I_1^2}{2\over qI_1s^2}+{E^2\over I_1^2}{2qI_1\over \sinh^2(qI_1s)}\right\}.
\end{eqnarray}
The integral is divergent, so we can compensate some EM field independent parts and find
\begin{eqnarray}
J_5^2&\Rightarrow&
-{\Omega\,qE^2\over8\pi^2 I_1}\int {\rm d}\,s~e^{-s\,m^2}\left[m^2\left({1\over\tanh(qI_1s)}-{1\over qI_1s}\right)+{1\over qI_1s^2}-{qI_1\over \sinh^2(qI_1s)}\right]\stackrel{m\rightarrow0}{=}-{\Omega |q|E^2\over4\pi^2I_1},
\end{eqnarray}
which is the same as that in Eq.(4) of the main text. Note that the EM field independent parts do not contribute to the axial current according to the vanishing of the chiral vortical effect in the $\mu,T\rightarrow0$ limit.

At last, the vector current can also be calculated as
\begin{eqnarray}
J^3&=&-{\Omega\over2}{\rm Tr}\;\gamma^3\hat{\cal S}({p})\gamma^4\gamma^1\gamma^3\hat{\cal S}({p})={\Omega\,qE\over2\pi I_1}\int{{\rm d}\,P_3{\rm d}\,P_4\over(2\pi)^2}\int {\rm d}\,s\;{\rm d}\,t~{e^{-(s+t)\left(m^2+P_3^2+P_4^2\right)}}\left(m^2+P_4^2-P_3^2\right).
\end{eqnarray}
As mentioned in the main text, we have to follow the dimensional regularization for the anomalous transport and find, in the limit $m\rightarrow0$,
\begin{eqnarray}
J^3={\Omega\,qE\over2\pi I_1}\int{{\rm d}\,P_3{\rm d}\,P_4\over(2\pi)^2}{P_4^2-P_3^2\over \left(P_3^2+P_4^2\right)^2}={\Omega}{qE\over4\pi}\!\lim_{q_0\rightarrow0}\!\lim_{q_3\rightarrow0}(-i\langle J^3(q)J^3(-q)\rangle_{2D}){=}{\Omega}{qE\over4\pi^2},
\end{eqnarray}
which reproduces that in Eq.(4) of the main text.